\documentstyle[12pt]{article}
\newcommand{\inv}{\left. g^0_A \right|_{\rm inv}}
\newcommand{\gam}{\mbox{${\gamma}^{\hspace{-2.5mm}
\raisebox{0.5mm}[0mm][0mm]{${\scriptstyle \leftrightarrow}$}}$}}
\newcommand{\Frac}[2]%
{{\textstyle \frac{\mbox{\footnotesize $#1$}\rule[-0.9mm]{0mm}{1mm}}%
{\mbox{\footnotesize $#2$}\rule{0mm}{3.1mm}}}}
\begin{document}
\begin{titlepage}
\vspace*{-12 mm}
\noindent
\begin{flushright}
hep-ph/9701213 
\end{flushright}
\vskip 20 mm
\begin{center}
{\large\bf Measuring anomalous ``spin'' in elastic $ep$ or $\nu p$\\[3mm]  
and deep inelastic $ep$ scattering}\\[16 mm]
{\bf S.D. Bass}$^{a,}$\footnote{sbass@pythia.itkp.uni-bonn.de},
{\bf R.J. Crewther}$^{b,c,}$\footnote{Leave of absence from the University 
of Adelaide ends February 5, 1997. \\
\makebox[1.8 em]{}rcrewthe@physics.adelaide.edu.au, crewther@butp.unibe.ch},
{\bf F.M. Steffens}$^{c,d,}$\footnote{fsteffen@finpe.if.usp.br},
{\bf A.W. Thomas}$^{c,}$\footnote{athomas@physics.adelaide.edu.au}
\\[10mm]   
$^a$ {\em Institut f\"{u}r Theoretische Kernphysik, 
Universit\"{a}t Bonn,\\
Nussallee 14--16, D-53115 Bonn, Germany}\\[5mm]
$^b$ {\em Institut f\"{u}r Theoretische Physik, Universit\"{a}t Bern,\\
Sidlerstrasse 5, CH-3012 Bern, Switzerland}\\[5mm]
$^c$ {\em Department of Physics and Mathematical Physics and\\ 
Centre for the Subatomic Structure of Matter,\\
University of Adelaide, Adelaide, SA 5005, Australia}\\[5mm]
$^d$ {\em Instituto de F\'{\i}sica da Universidade de S\~{a}o Paulo,\\
P.O. Box 66318, S\~{a}o Paulo 05315 970, SP, Brazil}
\end{center}
\vskip 12 mm
\begin{abstract}
\noindent
We obtain a general rule that the $O(1/\log m_h)$ 
term due to the current $\bar{h}\gamma_\mu\gamma_5h$ of a mass-$m_h$
quark $h$ in $f$-flavour theory is 
$-3\bar{g}^2_f(m_h)/\{2\pi^2(33-2f)\}$ times the flavour singlet current
of the residual $(f-1)$-flavour theory, where $\bar{g}_f$ is the
$f$-flavour running coupling constant in a mass-independent 
renormalization scheme. The rule is applied to the Ellis-Jaffe moment 
below and well above charm threshold, and to low-energy $Z^0$-exchange
amplitudes. The singlet axial charge of the proton common
to these experiments is both scale and gauge invariant, but is related
to the axial anomaly and the ``gluon spin'' by a non-perturbative 
renormalization factor.
\end{abstract}
\end{titlepage}
\renewcommand{\labelenumi}{(\alph{enumi})}
\renewcommand{\labelenumii}{(\roman{enumii})}
\newpage
With major experiments underway to determine the strange-quark
contributions to polarised deep inelastic scattering and low-energy
parity-violating elastic $ep$ and $\nu p$ scattering, it is 
vital to understand precisely what will be learnt. At first sight,
the answer seems clear: throw away contributions from heavy quarks,
including charm, and identify the remaining flavour singlet and
non-singlet contributions to axial charges of the proton in 
the various reactions. 

However the situation in QCD is far from obvious because of
renormalization effects in the singlet channel \cite{collins,kaplan,kod}
associated with the axial anomaly \cite{adler,bell}. In separate
studies of the deep inelastic \cite{mink} and elastic neutral current 
\cite{chet} processes mentioned above, it has been noted that these
effects give rise to a characteristic factor\footnote{Many earlier
papers, from \cite{collins} or \cite{kod} onwards, make implicit 
reference to this factor but then either omit it or keep only the 
leading order in the exponent. The lower limit of integration in 
equation (9) of \cite{larin} is missing; presumably $0$ was intended.
We follow the convention of \cite{kod} for $\gamma(g)$, i.e.\ the 
two-loop term is positive.}  
\begin{equation}
E(g) = \exp \int^g_0 \! dg'\, \gamma(g')/\beta(g')
\label{a}\end{equation}
multiplying the gauge-invariantly renormalized singlet axial-vector 
operator 
\begin{equation}
J^{GI}_{\mu5} = \left(\bar{u}\gamma_\mu\gamma_5u
                  + \bar{d}\gamma_\mu\gamma_5d
                  + \bar{s}\gamma_\mu\gamma_5s\right)_{GI}
\label{aa}\end{equation} 
in the residual theory (only three flavours $u,d,s$ if $c,b,t$ are 
considered heavy). Here $\beta(g)$ and $\gamma(g)$ are the 
Callan-Symanzik functions of the residual theory associated with the 
gluon coupling constant $g$ and the composite operator 
$J^{GI}_{\mu5}$. We note that the factor (\ref{a}) is 
non-perturbative: it involves integration over a finite range from 
$0$ to a value $g$ defined at a renormalization point $\mu$ well 
below the charm threshold.

Our aim is to pin down the relation between the axial proton charges 
which arise in these inelastic and elastic processes, while taking
full account of complications such as the factor (\ref{a}). This turns
out to be possible for leading terms in the heavy-quark expansion
\begin{equation}
m_t, m_b, m_c \longrightarrow \infty\ , \ \ m_t\gg m_b\gg m_c
\label{b}\end{equation}
and includes a simple rule for the $O(1/\log m_h)$ correction as each
quark $h=t,b,$ or $c$ is made infinitely heavy. For the inelastic
process, two limits are considered:
\begin{enumerate} 
\item Asymptotia below charm threshold, defined by
\begin{equation}
m_t \gg m_b \gg m_c \gg Q \gg \mu 
\label{below}\end{equation}
where $\mu$ is the renormalization subtraction point,
i.e. all heavy quark limits, including $m_c \rightarrow \infty$, 
are taken {\em before\/} the limit $Q = \sqrt{- q^2} \rightarrow \infty$ 
of large space-like momentum transfer $q_\mu$.
\item At the end of the paper, asymptotia well above charm threshold,
but below bottom threshold, {\em viz.\/}
\begin{equation}
m_t \gg m_b \gg Q \gg m_c \gg \mu 
\label{above}\end{equation}
\end{enumerate}

We begin with inelastic polarized $ep$ scattering in the limit
(\ref{below}), the residual theory being QCD with three flavours 
$u,d,s$. According to (\ref{below}), asymptotia for a structure
function $W$ is the deep inelastic limit $Q \rightarrow \infty$ taken 
{\em after\/} the limit (\ref{b}),
\begin{equation}
W(x,\infty) = \lim_{Q \rightarrow \infty}\lim_{\rm eq\ (\ref{b})}
               \left. W(x,Q^2)\right|_{\rm 6\ flavours}
\label{c}\end{equation}
where $x = Q^2/2p.q$ is Bjorken's variable. This means that
the relevant deep inelastic data from which a three-flavour
singlet charge is to be extracted should be restricted to kinematic
regions below the charm threshold ($\Lambda_c D$ production):
\begin{equation}
Q^2(1-x)/x\ <\ (m_D + m_{\Lambda_c})^2 - m_p^2
\end{equation}
Small violations of this condition may be acceptable, provided that 
all charm threshold effects are isolated \cite{donn} and removed.

Let $\bar{g}=\bar{g}(Q)$ be the effective coupling constant of the
three-flavour theory,
\begin{equation}
\log(Q/\mu) = \int_g^{\bar{g}} \! dg'/\beta(g')
\end{equation}
and let $\inv$ be the axial charge of the operator
\cite{mink,larin}
\begin{equation}
{\cal S}_\mu(x) = E(g)J^{GI}_{\mu5}(x)
\label{f}\end{equation}
for proton states $|p,s\rangle$ with momentum $p_\mu$ and spin 
$s_\mu$:
\begin{equation}
\langle p,s|{\cal S}_\mu(0)|p,s\rangle = 2m_p s_\mu \inv
\label{e}\end{equation}
Then, as $Q$ tends to $\infty$, the $SU(3)$ flavour singlet part of 
the Ellis-Jaffe moment \cite{ellis} is given by 
\cite{kod,mink,larin,alta,grun}
\begin{equation}
\int^1_0\! dx \left. g_1(x,Q^2)\right|_{\rm singlet}\ 
\sim\ \Frac{1}{9}\!\inv
\Bigl\{1 + \sum_{\ell\geq 1} c_\ell\,\bar{g}^{2\ell}(Q)\Bigr\}
\label{d}\end{equation}
where the coefficient $c_\ell$ is calculable in $\ell$-loop 
perturbation theory \cite{larin}.

The factor $E(g)$ in (\ref{f}) arises naturally from the coefficient
function of $J^{GI}_{\mu5}$ in the product of electromagnetic currents
$J_\alpha(x)J_\beta(0)$ at short distances $x_\mu \sim 0$. It
compensates for the scale dependence of $J^{GI}_{\mu5}$ caused by the
anomaly in its divergence\footnote{As Adler noted in his original 
paper \cite{adler}, the anomaly gives rise to a multiplicative factor
$Z_5$ which makes $J^{GI}_{\mu5}$ finite at two-loop level and
beyond. The corresponding $\gamma(g)$ function was introduced by 
K\"{o}berle and Nielsen \cite{koeb} in quantum electrodynamics (before
asymptotic freedom). The extension to QCD was discussed in the review 
\cite{rjc} and independently in \cite{collins}. There followed
Kodaira's well-known application \cite{kod} to polarised deep
inelastic scattering.}, 
\begin{equation}
\partial^\mu J^{GI}_{\mu5}
= 2f\partial^\mu K_\mu + \sum_{i=1}^{f} 2im_i \bar{q}_i\gamma_5 q_i
\end{equation}
where $K_\mu$ is a renormalized version of the gluonic Chern-Simons 
current, and the number of flavours $f$ is $3$. 

A consequence of the renormalization-scale invariance of 
${\cal S}_\mu$ is that its spatial components have operator charges
\begin{equation}
S_i(t) = \int\! d^3x\, {\cal S}^i(t,{\bf x})
\end{equation}
which satisfy an equal-time algebra \cite{mink}
\begin{equation}
[S_i(t),S_j(t)] = i\varepsilon_{ijk}S_k(t)
\end{equation}
characteristic of spin operators.
Unfortunately, the time dependence of the ``spin'' $S_i$
implies that energy eigenstates such as $|p,s\rangle$ are complicated 
mixtures of its representations. Note that a proposal 
\cite{beg} to obtain the $U(6)\times U(6)$ algebra \cite{fgmz} 
in QCD works only if the singlet axial-vector operator is identified
as ${\cal S}_\mu$. 

As noted above, the factor (\ref{a}) is not perturbative, and so should 
{\em not\/} be approximated by retaining just the lowest-order terms 
in $\beta$ and $\gamma$. We also note that, although the divergence of
${\cal S}_\mu$ is
\begin{equation}
\partial^\mu{\cal S}_\mu = 6E(g)\partial^\mu K_\mu + \mbox{mass terms}
\end{equation}
the quantity $E(g)\partial^\mu K_\mu$ is not scale invariant
(contrary to \cite{grun}) because $K_\mu$ is not multiplicatively
renormalized%
\footnote{This point has been rediscovered several
times. Again, the essence of it is in Adler's original paper 
\cite{adler}, being connected with non-renormalization of the
anomalous constant. The minor changes needed for QCD were discussed
in the review \cite{rjc}. See the reviews\cite{alta,cheng} 
for recent discussions of this point in the context of proton 
``spin''.}: 
its counterterms are all proportional to $J^{GI}_{\mu5}$. As a 
result, in the relation between the observed singlet charge $\inv$,
the ``gluon spin'' $\Delta g$ \cite{early,efremov} as interpreted in 
the light-front gauge \cite{jaffe}, and  the scale invariant 
contribution $\Delta\Sigma_{\rm cons}$ from the gauge dependent, 
partially conserved chiral current, $E(g)$ appears as an
uncontrolled theoretical factor:%
\footnote{This reflects the well-known fact that the ratio of the 
moment (\ref{d}) at $Q = \infty$ to that at low $Q$ is 
non-perturbative. Our point here is that we will be able to relate 
deep inelastic scattering to elastic $Z^0$ exchange {\em despite\/} 
the presence of such a factor.}
\begin{equation}
\inv = E(g)\Bigl(\Delta\Sigma_{\rm cons} 
- \frac{3g^2}{8\pi^2}\Delta g\Bigr)
\end{equation} 
The scale dependence of $\Delta g$ shows that the {\em canonical\/}
basis for supposing it to be gluonic spin is destroyed by 
renormalization. We emphasize that the results of this paper do not 
depend on assumptions about $E(g)$ or $\Delta g$. 

So, returning to (\ref{d}), we arrive at a conclusion which 
is now essentially conventional: the three-flavour 
singlet Ellis-Jaffe moment (\ref{d}) at $Q \sim \infty$ tends 
to a singlet axial charge $\left. g^0_A\right|_{\rm inv}$ which is 
both gauge and scale invariant, as any measurable quantity should 
be. Its current experimental value is \cite{expt,badelek}
\begin{equation}
\left. g^0_A \right|_{\rm inv} = 0.28 \pm 0.07 
\label{fa}\end{equation}
Other notations for this quantity are $\Sigma_{\rm inv}$ \cite{larin}
and $\Delta\Sigma(\infty)$ \cite{alta}.

Now let us consider whether \cite{bass} this deep-inelastic version of
the singlet axial charge is relevant for elastic $Z^0$-exchange
reactions. Once again, we will consider the heavy-quark limit 
(\ref{b}), in this case, applied to the axial-vector neutral current
\begin{equation}
J_{\mu5}^Z\ 
=\ \Frac{1}{2} \biggl\{\,\sum_{q=u,c,t} - \sum_{q=d,s,b}\,\biggr\}\:
        \bar{q}\gamma_\mu\gamma_5q
\label{g}\end{equation}
coupled to the proton. 

Originally, Collins et al.\ \cite{collins} estimated residual effects 
from anomaly-cancelling pairs $(t,b)$, $(c,s)$, \ldots in (\ref{g}). 
Later Kaplan and Manohar \cite{kaplan} considered how to extend the 
method to the more realistic case where $c$ is heavy and $s$ is
light. The procedure adopted in these papers was to set a 
renormalization scale $\mu_{\rm int}$ between each quark made 
heavy and the next in line. This approach belongs 
to the class of mass-dependent renormalization schemes in which
decoupling theorems \cite{sym} are most easily proven. Comparisons 
with mass-independent schemes, where decoupling is not manifest, were 
the subject of much discussion many years ago \cite{witten,wein}.
 
However, mass-independent schemes are preferable when one wants 
to develop explicit asymptotic expansions in $1/m_h$. For example, 
Chetyrkin and K\"{u}hn \cite{chet}, commenting on the result of 
\cite{collins}, observed that the renormalization prescription of 
amplitudes in the residual theory produced by a heavy-quark limit 
is hard to determine in a mass-dependent renormalization scheme. 
It is much simpler to have asymptotic results depending on just 
one scale parameter $\mu$ and not on the residual quark masses. 

So we choose to work in a mass-independent scheme, choosing $\mu$ 
such that it is suitable for the final residual theory with just 
three flavours, and relying on the renormalization group to keep track 
of all factors. This means that we consider $\mu$ to be well below 
$m_c$, as in the case of deep inelastic scattering. 

Let ${\cal A}_6$ be an amplitude in the full six-flavour theory. 
Its expansion in the inverse top mass $m_t^{-1}$ looks like an 
operator-product expansion in momentum space, but with $Q$ replaced 
by $m_t$. The result involves six-flavour coefficient functions 
${\cal C}_{n6}(m_t)$ which multiply five-flavour amplitudes 
${\cal A}_{n5}$: 
\begin{equation}
{\cal A}_6 \sim \sum_n {\cal C}_{n6}(m_t) {\cal A}_{n5}\ , 
\ \ m_t \rightarrow \infty
\label{ga}\end{equation}
Since all ${\cal C}_{n6}(m_t)$ are obtained at $m_t\sim\infty$, they 
correspond to sub-amplitudes in which all light-quark ($q\not= t$) and 
gluon propagators carry loop momenta $O(m_t)$. This means that,
provided the renormalization scheme is mass-independent, these
coefficient functions do not depend on $m_b$ or lighter quark masses. 

Consequently, the heavy-$b$ expansion can be obtained by expanding 
each five-flavour amplitude ${\cal A}_{n5}$ in (\ref{ga}) in terms 
of five-flavour coefficient functions ${\cal C}_{nn'5}$ and 
four-flavour amplitudes ${\cal A}_{nn'4}$,
\begin{equation}
{\cal A}_{n5} = \sum_{n'} {\cal C}_{nn'5}(m_b) {\cal A}_{nn'4}\ ,
\ \ m_b \rightarrow \infty
\end{equation}
and substituting back into (\ref{ga}). The limit sequence can be
continued by then expanding ${\cal A}_{nn'4}$ in $m_c^{-1}$, but 
no further, since $c$ is the lightest quark which can conceivably 
be treated as heavy. 

The full expansion of the original amplitude (\ref{ga}) is thus
\begin{equation}
{\cal A}_6 \sim \sum_{nn'n''}{\cal C}_{n6}(m_t){\cal C}_{nn'5}(m_b) 
{\cal C}_{nn'n''4}(m_c) {\cal A}_{nn'n''3}\ , 
\ \ m_t \gg m_b \gg m_c \gg \mu
\label{h}\end{equation}
It involves non-perturbative three-flavour amplitudes 
${\cal A}_{nn'n''3}$ which we would like to relate to quantities
such as $\inv$.

Let the six-flavour amplitude (\ref{ga}) be the proton matrix element
of the neutral current (\ref{g}):
\begin{equation}
{\cal A}_6 = \langle p,s|J^Z_{\mu5}|p,s\rangle
\label{gb}\end{equation}
It is well known that $J^Z_{\mu5}$ has a soft divergence and so 
is renormalization-scale invariant in the full six-flavour theory. 
Consequently, the amplitude (\ref{gb}) is scale invariant, as is
required for a measurable quantity. For any heavy-quark limit, it 
must remain measurable and scale invariant in the {\em residual\/} 
theory with fewer flavours. 

First we consider terms in (\ref{h}) which survive the limiting 
procedure; logarithmic corrections will be considered later. The
residual theory has three flavours, so flavour SU(3) should be used
to classify operators in the expansion of $J^Z_{\mu5}$. The
only possibilities are the octet operators
\begin{equation}
J^a_{\mu5} = \Frac{1}{2}\bar{q}\gamma_\mu\gamma_5\lambda^a q
\end{equation}
and the singlet operator (\ref{aa}), or better, the scale invariant 
operator ${\cal S}_\mu$ defined in (\ref{f}), since we know that the
answer must be scale invariant. Clearly, the terms $q=u,d$ in 
(\ref{g}) give $J^3_{\mu5}$, the term $q=s$ can be split into
$-J^8_{\mu5}/\surd{3}$ and something proportional to ${\cal S}_\mu$,
while the terms $q=t,b,c$ contribute to ${\cal S}_\mu$ only:
\begin{equation}
J^Z_{\mu5} \longrightarrow 
   \mbox{octet} + \{\mbox{constant}\}{\cal S}_\mu
\label{j}\end{equation}
By construction, the constant of proportionality in (\ref{j}) is 
scale invariant. It is also quark-mass independent, since we use a 
mass-independent renormalization scheme, so it must be 
independent of the gluon coupling constant $g$. Therefore we can 
evaluate it by considering the limit $g\rightarrow 0$ for 
$J^Z_{\mu5}$ coupled to off-shell $u,d,s$ quarks and noting 
that $E(0)=1$: 
\begin{equation}
\mbox{constant} = -\Frac{1}{6}
\label{ja}\end{equation}
We conclude that, in the strict heavy-quark limit (\ref{b}), the 
neutral current is given by
\begin{equation}
J^Z_{\mu5} \longrightarrow 
J^3_{\mu5} + \Frac{1}{\surd{3}}J^8_{\mu5} - \Frac{1}{6}{\cal S}_\mu
\label{jb}\end{equation} 

Consequently, there is a precise relation between the three-flavour 
Ellis-Jaffe moment (\ref{d}) and the neutral current amplitude 
(\ref{gb}) in the limit (\ref{below}),
\begin{equation}
\langle p,s|J^Z_{\mu5}|p,s\rangle  \longrightarrow 2m_p s_\mu 
\left( g_A^{(3)} + \Frac{1}{\surd{3}}g_A^{(8)} - \Frac{1}{6}\!\inv\right)
\label{jd}\end{equation}
where $g_A^{(a)}$ are the usual octet axial charges
\begin{equation}
\langle p,s|J^a_{\mu5}|p,s\rangle = 2m_p s_\mu g_A^{(a)}\ ,\ \ a=3,8
\end{equation}
(so $g_A/g_V$ for neutron decay equals $2g^3_A$).
Note that (\ref{jd}) holds despite the presence of $E(g)$. The matching
condition (\ref{ja}) at $g=0$ is crucial in arriving at this otherwise
conventional conclusion. It can be made to look even more
conventional,
\begin{equation}
\langle p,s|J^Z_{\mu5}|p,s\rangle  \longrightarrow (2m_p s_\mu)\,
  \Frac{1}{2}\left(\Delta u_{\rm inv} - \Delta d_{\rm inv} 
    - \Delta s_{\rm inv}\right)
\end{equation}
if we define scale invariant axial charges $\Delta q_{\rm inv}$ 
for $q=u,d,s$ via the prescription
\begin{equation}
\left(\begin{array}{c} 
\Delta u_{\rm inv} \vphantom{g_A^{(3)}} \\ 
\Delta d_{\rm inv} \vphantom{g_A^{(8)}} \\ 
\Delta s_{\rm inv} \vphantom{\inv} \\  
\end{array}\right)\
=\
\frac{1}{3}\left(\begin{array}{ccc} 
3 & \surd{3} & 1 \vphantom{g_A^{(3)}} \\  
-3 & \surd{3} & 1 \vphantom{g_A^{(8)}} \\  
0 & -2\surd{3} & 1 \vphantom{\inv}  
\end{array}\right)
\left(\begin{array}{c} g_A^{(3)} 
\\ g_A^{(8)} 
\\ \inv
\end{array}\right)
\label{je}\end{equation}
Apart from its strong dependence on the heavy-quark limit, the
relation between (\ref{d}) and (\ref{jd}) is an isoscalar version
of Bjorken's sum rule \cite{bj}, 

The next step is to determine the leading $O(1/\log m_h)$ corrections
to (\ref{jd}). 

Let us retain only the leading power in the general expansion 
(\ref{ga}). If we arrange that each independent
five-flavour amplitude ${\cal A}_5$ remaining on the right-hand
side is scale invariant, the result necessarily consists of one
or more expressions of the form
\begin{equation}
{\cal A}_6 \sim 
{\cal A}_5\sum_{\ell\geq 0} C_{\ell6}\,\bar{g}_6^{2\ell}(m_t) + \ldots
\end{equation}
where $\bar{g}_6$ is the effective gluon coupling in the six-flavour
theory and the coefficients $C_{\ell6}$ can be calculated
perturbatively. Proceeding as before, we find that the leading power
of (\ref{h}) consists of one or more terms
\begin{equation}
{\cal A}_6\ \sim\ 
{\cal A}_3\sum_{\ell,\ell',\ell''\geq 0} C_{\ell6}C_{\ell'5}C_{\ell''4}
\,\bar{g}_6^{2\ell}(m_t)\,\bar{g}_5^{2\ell'}(m_b)\,\bar{g}_4^{2\ell''}(m_c) 
+ \ldots
\label{k}\end{equation}
involving the effective gluon couplings $\bar{g}_f$ for $f$ flavours
as functions of $m_t,m_b,m_c$ for $f=6,5,4$ respectively. (In this
notation, the effective coupling $\bar{g}$ in (\ref{d}) would be written
$\bar{g}_3(Q)$.)

Suppose that, at a given stage of the calculation, $f$ flavours
remain, and $h$ is the next quark to be made heavy. As noted by
Collins et al.\ \cite{collins}, the $O(1/\log m_h)$ correction
arises because the current $\bar{h}\gamma_\mu\gamma_5 h$ in $J^Z_{\mu5}$
can be part of a triangle sub-diagram formed from $h$-quark
propagators. For $m_h$ large, $\bar{h}\gamma_\mu\gamma_5 h$ mixes with
the operator $\sum_{q\not= h}\bar{q}\gamma_\mu\gamma_5 q$.
We need the scale-invariant versions of these currents contained in the 
column vector
\begin{equation}
{\cal V}_\mu\ 
=\ \left(\begin{array}{c} 
\bar{h}\gamma_\mu\gamma_5 h \\
\sum_{q\not= h}\bar{q}\gamma_\mu\gamma_5 q
\end{array}\right)_{\rm inv}\
=\ \left(\begin{array}{cc} f^{-1} & -1 \\ 1 - f^{-1} & 1 
\end{array}\right)
\left(\begin{array}{c} 
\{{\cal S}_\mu\}_f \\ \{Y_{\mu5}\}_f
\end{array}\right)
\end{equation}
where the equations
\begin{eqnarray}
\{{\cal S}_\mu\}_f &=&
\{E(g)\}_f \Bigl(\sum_{i=1}^f\bar{q}_i\gamma_\mu\gamma_5 q_i\Bigr)_{GI} \\
\{Y_{\mu5}\}_f 
&=& f^{-1}\sum_{i=1}^f\bar{q}_i\gamma_\mu\gamma_5 q_i
    - \bar{h}\gamma_\mu\gamma_5 h
\end{eqnarray}
define $f$-flavour analogues of (\ref{f}) and the axial-vector
hypercharge current. Note that the currents in ${\cal V}_\mu$ are 
normalized such that at $g=0$, they reduce to the usual free-field 
expressions.

Let $\gam (g)$ be the matrix Callan-Symanzik function controlling the
mixing of ${\cal V}_\mu$ components as $m_h$ becomes large. 
At $m_h=\infty$, the lower component of ${\cal V}_\mu$ tends to 
$\{{\cal S}_\mu\}_{f-1}$, which is the analogue of (34) in
$(f-1)$-flavour theory. Therefore the entire leading power%
\footnote{Strictly speaking, we should replace the renormalized 
parameter $m_h$  by the scale invariant quantity 
$\kappa m_h \{\exp -\int^g_0 \gamma_m(g')/\beta(g')\}_f$ in all asymptotic
equations, where $\gamma_m(g)$ is the Callan-Symanzik function for
mass renormalization in the $f$-flavour theory, and $\kappa$ is a
number to be determined from the finite part of the two-loop calculation
at $m_h=\infty$. This does not affect $O(1/\log m_h)$ corrections.}
is given by
\begin{equation}
{\cal V}_\mu =
{\cal P}\biggl\{\exp\int^{\bar{g}(m_h)}_0 \!\! dg'\,
\frac{\gam (g')}{\beta(g')}\biggr\}_f
\left(\begin{array}{c} 
\{\bar{h}\gamma_\mu\gamma_5 h\}_{\rm inv}  \\
\{{\cal S}_\mu\}_{f-1}
\end{array}\right)_{m_h = \infty} + O(m^{-1}_h)
\label{m}\end{equation}
where ${\cal P}$ denotes ordering of the exponential. 

Our interest is in the upper off-diagonal term $g^4/16\pi^4 + O(g^6)$ 
in $\gam$ combined with  $-g^3(33-2f)/48\pi^2 + O(g^5)$ for the 
$f$-flavour $\beta$ function (for three colours). Substituting these 
expressions in (\ref{m}), we deduce our most interesting result, 
{\em viz.\/} the rule 
\begin{equation}
\{\bar{h}\gamma_\mu\gamma_5 h\}_{\rm inv}
= - \frac{3\bar{g}^2_f(m_h)}{2\pi^2(33-2f)} 
\{{\cal S}_\mu\}_{f-1} 
+ O\Bigl(\{\log (m_h)\}^{-2}\Bigr)
\label{ma}\end{equation}
for the leading $O(1/\log m_h)$ correction. This analytic result
is the first of a series of perturbative calculations which can 
now be done to determine various coefficients in (\ref{k}). The
techniques resemble those used to obtain $O(1/\log Q)$ expansions 
such as (\ref{d}). 

We should note that Chetyrkin and K\"{u}hn \cite{chet} come very close
to stating this rule, but they differ on a detail in the flavour 
dependence. This becomes clear when successive heavy-quark limits
are considered. Following \cite{collins}, they considered
$m_t\rightarrow\infty$ and then $m_b\rightarrow\infty$ 
for the difference 
$\bar{t}\gamma_\mu\gamma_5 t - \bar{b}\gamma_\mu\gamma_5 b$ 
and found a {\em five\/}-flavour expression 
\[ \Frac{3}{23}\Bigl(\bar{g}^2_5(m_b) - \bar{g}^2_5(m_t)\Bigr)/2\pi^2  \]
for the coefficient of the four-flavour singlet current. Our rule
requires a {\em six\/}-flavour dependence for the $O(1/\log m_t)$
term, $f=5$ being relevant only for the $O(1/\log m_b)$ term:
\begin{equation}
\bar{t}\gamma_\mu\gamma_5 t - \bar{b}\gamma_\mu\gamma_5 b
\sim \Bigl(\Frac{3}{23}\bar{g}^2_5(m_b) 
 - \Frac{1}{7}\bar{g}^2_6(m_t)\Bigr)\Frac{1}{2\pi^2}
\{{\cal S}_\mu\}_4
\label{mb}\end{equation}

It is not clear whether the result of Collins et al.\ \cite{collins} 
is consistent with (\ref{mb}) or not. Their formula contains only 
one running coupling $\bar{g}$ evaluated at both $m_t$ and $m_b$, 
so the flavour dependence does not appear to be the same. However, 
their renormalization scheme is mass dependent (unlike ours),
producing amplitudes in the residual theory with renormalization
prescriptions which are hard to specify \cite{chet}.  Comparisons
with the work of Kaplan and Manohar \cite{kaplan,manohar}, in which 
the mass-dependent approach of \cite{collins} is taken much further,
are correspondingly more difficult to make.

In the limit (\ref{b}), the rule (\ref{ma}) applied to the neutral
current implies
\begin{equation}
J^Z_{\mu5} = J^3_{\mu5} + \Frac{1}{\surd{3}}J^8_{\mu5}
               - {\cal C}(m_h){\cal S}_\mu + O(1/m_h)
\label{mc}\end{equation} 
where the coefficient ${\cal C}$ (including the factor 
$\frac{1}{2}$ in (\ref{g})) is
\begin{equation}
{\cal C} = \Frac{1}{6} + \Frac{3}{4\pi^2}\Bigl( 
\Frac{1}{21}\bar{g}^2_6(m_t) - \Frac{1}{23}\bar{g}^2_5(m_b)
+\Frac{1}{25}\bar{g}^2_4(m_c)\Bigr) + O\Bigl(\{\log(m_h)\}^{-2}\Bigr)
\end{equation} 
This coefficient fixes the first logarithmic correction for the 
amplitude (\ref{jd}):
\begin{equation}
\langle p,s|J^Z_{\mu5}|p,s\rangle = 2m_p s_\mu 
\left( g_A^{(3)} + \Frac{1}{\surd{3}}g_A^{(8)} 
- {\cal C}(m_h)\!\inv\right) + O(1/m_h)
\label{jf}\end{equation}
As in (\ref{ja}), normalizations can be checked via matching conditions at
$g\sim 0$ at any point in the derivation.

A numerical estimate of the corrections in ${\cal C}$ can be obtained
by substituting 
\begin{equation}
\bar{g}^2_f \sim \frac{24\pi^2}{33-2f}\{\log(m_h/\mu)\}^{-1}
\end{equation} 
with the scale $\mu$ chosen well below $m_c$ but in the range of a few to
several hundred MeV characteristic of low-energy strong interactions.
Let us take $\mu = 0.5$ GeV. Then, listing contributions in the order
$h=t,b,c$, we find
\begin{equation}
{\cal C}\ \approx\ 
\Frac{1}{6}(1 + 0.04 - 0.09 + 0.15)\ =\ \Frac{1}{6}(1.11)
\end{equation}
The correction is almost three times that of \cite{kaplan} and
opposite in sign. It corrects $\Delta s_{\rm inv}$ by an amount
$0.01$ compared with the experimental value 
\cite{expt,badelek} $-0.11 \pm 0.03$ for inelastic processes. 

The rule (\ref{ma}) can also be used to compare the Ellis-Jaffe
moment (\ref{d}) below charm threshold, as defined by the limiting
procedure (\ref{below}), and the corresponding quantity 
$\int^1_0 dx g^c_1(x,Q^2)$ well above threshold, where asymptotia 
is defined by the condition (\ref{above}).
Note that $c$ is still considered heavy, not light (since flavour
$SU(4)$ symmetry is a very bad approximation compared with 
$m_c \gg \mu$). Going from (\ref{below}) to (\ref{above}), we have 
to replace the three-flavour singlet operator ${\cal S}_\mu$ by the 
four-flavour expression 
\begin{equation}
\Big\{\sum_{q\not= c}\bar{q}\gamma_\mu\gamma_5 q 
+ 2\bar{c}\gamma_\mu\gamma_5 c\Bigr\}_{\rm inv}
\end{equation} 
Consequently, as $m_c$ becomes large, the rule (\ref{ma}) predicts
\begin{eqnarray}
\int^1_0 \! dx\,\Bigl(g^c_1(x,\infty) - g_1(x,\infty)\Bigr)
\sim -\Frac{3}{25\pi^2}\bar{g}^2_4(m_c)\,\Frac{1}{9}\!\inv
\approx \Frac{1}{9}\!\inv (-0.10)
\label{jg}\end{eqnarray}
where $\inv$ is the three-flavour singlet axial charge (\ref{e})
appearing in (\ref{d}) and (\ref{jf}). The minus sign in (\ref{jg})
indicates that charm production is slightly enhanced when the
beam and target spins are aligned.

A previous estimate \cite{AL} of (\ref{jg}) was obtained by evolving
the moment through the charm threshold. This involves a specific model
for threshold behaviour for scale-dependent amplitudes in a 
mass-dependent renormalization prescription, so a theoretical or
numerical comparison with our result is not obvious. Throughout, we
avoid thresholds and use a mass-independent prescription; that leads
to explicit results for $O(1/\log m_h)$ corrections.
\vskip 4mm
\noindent
{\large\bf Acknowledgements}
\vskip 2mm

Support from the Alexander von Humboldt Foundation (SDB and AWT), 
the Australian Research Council (AWT), and CAPES Brazil (FMS)
is gratefully acknowledged. FMS and AWT thank the Institute
for Nuclear Theory at the University of Washington for its hospitality
and the U.S. Department of Energy for partial support at the beginning
of this work. RJC thanks Peter Minkowski for helpful discussions, and 
Dieter Sch\"{u}tte for his hospitality at the Institute for
Theoretical Nuclear Physics, University of Bonn, where the manuscript
was completed.


\begin{thebibliography}{99}
\bibitem{collins} J. Collins, F. Wilczek and A. Zee, Phys. Rev.
D 18 (1978) 242.
\bibitem{kaplan} D.B. Kaplan and A.V. Manohar, Nucl. Phys. B 310
(1988) 527.
\bibitem{kod} J. Kodaira, Nucl. Phys. B 165 (1980) 129.
\bibitem{adler} S.L. Adler, Phys. Rev. 177 (1969) 2426.
\bibitem{bell} J.S. Bell and R. Jackiw, Nuovo Cimento 60 A (1969)
47.
\bibitem{mink} P. Minkowski, {\em in\/} Proc. Workshop on {\em
Effective Field Theories of the Standard Model\/}, Dobog\'{o}k\~{o},
Hungary 1991, ed. U.-G. Meissner (World Scientific, Singapore, 1992).
\bibitem{chet} K.G. Chetyrkin and J.H. K\"{u}hn, Z. Phys. C 60 (1993) 
497.
\bibitem{larin} S.A. Larin, Phys. Lett. B 334 (1994) 192.
\bibitem{donn} A. Donnachie and P.V. Landshoff, Z. Phys. C 61 (1994)
139.
\bibitem{ellis} J. Ellis and R.L. Jaffe, Phys. Rev. D 9 (1974) 1444;
(E) D 10 (1974) 1669; \\
R.L. Jaffe, Phys. Lett. B 193 (1987) 101.
\bibitem{alta} G. Altarelli and G. Ridolfi, Nucl. Phys. B (Proc. 
Suppl.) 39\,B,C (1995) 106.
\bibitem{grun} G. Grunberg, Phys. Lett. B 388 (1996) 173.
\bibitem{cheng} H.-Y. Cheng, Int. J. Mod. Phys. A 11 (1996) 5109.
\bibitem{koeb} R. K\"{o}berle and N.K. Nielsen, Phys. Rev. D 8
(1973) 660;\\
R.J. Crewther, S.-S. Shei and T.-M. Yan, Phys. Rev. D 8 (1973) 3396;\\
R.J. Crewther and N.K. Nielsen, Nucl. Phys. B 87 (1975) 52.
\bibitem{rjc} R.J. Crewther, Effects of Topological Charge in Gauge
Theories, in Facts and Prospects of Gauge Theories, Schladming,
Austria, February 1978, ed. P. Urban, Acta Physica Austriaca Suppl. 
19 (1978) 47, section 7.
\bibitem{beg} M.A.B. B\'{e}g, Phys. Rev. D 11 (1975) 1165.
\bibitem{fgmz} R.P. Feynman, M. Gell-Mann, and G. Zweig, Phys. Rev.
Lett. 13 (1964) 678.
\bibitem{early} C.S. Lam and B.-A. Li, Phys. Rev. D 25 (1982) 683;\\
P. Ratcliffe, Nucl. Phys. B 223 (1983) 45.
\bibitem{efremov} A.V. Efremov and O.V. Teryaev, JINR Report
E2--88--287 (1988), and in Proceedings of the International Hadron
Symposium, Bechyn\v{e} 1988, eds.\ J. Fischer et al.\ 
(Czechoslovakian Academy of Science, Prague, 1989) p. 302;\\
G. Altarelli and G.G. Ross, Phys. Lett. B 212 (1988) 391;\\
R.D. Carlitz, J.C. Collins, and A.H. Mueller, Phys. Lett. B 214
(1988) 229.
\bibitem{jaffe} R.L. Jaffe and A. Manohar, Nucl. Phys. B 337 (1990)
509;\\
R.L. Jaffe, Phys. Lett. B 365 (1996) 359.
\bibitem{expt} EMC Collaboration, J Ashman et al., Phys. Lett. B 206
(1988) 364; Nucl. Phys. B 328 (1989) 1;\\
SMC Collaboration, D. Adams et al., Phys. Lett. B 329 (1994) 399;
(E) B 339 (1994) 332\\
E143 Collaboration, K. Abe et al., Phys. Rev. Lett. 74 (1995) 346. 
\bibitem{badelek} B. Bade{\l}ek, CRAD96 conference, Cracow, August
1996 (hep-ph/9612274).
\bibitem{bass} S.D. Bass and A.W. Thomas, Phys. Lett. B 293 (1992)
457.
\bibitem{bj} J.D. Bjorken, Phys. Rev. 148 (1966) 1467; Phys. Rev.
D 1 (1970) 1376.
\bibitem{sym} K. Symanzik, Commun. Math. Phys. 34 (1973) 7;\\
T. Appelquist and J. Carazzone, Phys. Rev. D 11 (1975) 2856. 
\bibitem{witten} E. Witten, Nucl. Phys. B 104 (1976) 445.
\bibitem{wein} S. Weinberg, Phys. Lett. B 91 (1980) 51;\\
B.A. Ovrut and H.J. Schnitzer, Phys. Rev. D 21 (1980) 3369; Nucl.
Phys. B 179 (1981) 381;\\ 
W. Bernreuther and W. Wetzel, B 197 (1982) 228;\\
W. Bernreuther, Ann. Phys. (N.Y.) 151 (1983) 127;\\
W. Marciano, Phys. Rev. D 29 (1984) 580.
\bibitem{manohar} A.V. Manohar, Phys. Lett. B 242 (1990) 94.
\bibitem{AL} G. Altarelli and B. Lampe, Z. Phys. C 47 (1990) 315.

\end{thebibliography}
\end{document}